\documentclass{article}
\usepackage{mrs2005,epsfig}
\def\R23{\mbox{$\rm R_{23}$}}

\def\Hb{\mbox{${\rm H}{\beta}$}}
\def\Ha{\mbox{${\rm H}{\alpha}$}}
\def\OIIIa{\mbox{${\rm [O\,III]\,}{\lambda\,5007}$}}

\def\OII{\mbox{${\rm [O\,II]\,}{\lambda\,3727}$}}

\def\NII{\mbox{${\rm [N\,II]\,}{\lambda\,6584}$}}

\begin{document} 
\title{Oxygen Gas Abundances at $0.4<z<1.5$: Implications for the Chemical Evolution History of Galaxies}

\author{C. Maier, S.J. Lilly, C.M. Carollo}
\affil{Department of Physics, Swiss Federal Institute of Technology (ETH Z\"urich), ETH H\"onggerberg, CH-8093, Z\"urich, Switzerland}

\begin{abstract}
 We report VLT-ISAAC   and Keck-NIRSPEC near-infrared spectroscopy
 for a sample of 30 $0.47<z<0.92$ CFRS galaxies and five [OII]-selected,
  $M_{B,AB}<-21.5$, $z\sim1.4$ galaxies. We have
  measured \Ha\, and \NII\, line fluxes for  the CFRS galaxies
  which have \OII, \Hb\, and \OIIIa\, line fluxes available from 
 optical spectroscopy. For the $z\sim1.4$ objects we  measured \Hb\, and \OIIIa\,
  emission line fluxes from J-band spectra, and \Ha\, line fluxes plus upper
  limits for \NII\, fluxes from H-band spectra.
  We derive the extinction and oxygen abundances for the  sample using a method  based on
  a set of ionisation parameter and oxygen abundance diagnostics, 
  simultaneously fitting the [OII], \Hb,
  [OIII], \Ha, and [NII] line fluxes.
Our most salient conclusions are:
a) the  source of gas
ionisation in the 30  CFRS and in all $z\sim1.4$ galaxies is not due
to AGN activity;
b) about one third of the $0.47<z<0.92$ CFRS
  galaxies in our sample have substantially lower metallicities than local galaxies with similar
  luminosities and star formation rates;
c) comparison with a chemical evolution model indicates that these low
  metallicity galaxies
 are unlikely to be the progenitors of metal-poor dwarf galaxies at
 $z\sim0$, but more likely the progenitors of massive spirals;
d) the $z\sim1.4$
  galaxies are characterized by
  the high [OIII]/[OII] line ratios, low extinction and low metallicity that
  are typical of lower luminosity CADIS galaxies at $0.4<z<0.7$, and of
  more luminous Lyman Break
  Galaxies at $z\sim3.1$, but not seen in  CFRS galaxies at $0.4<z<1.0$; 
e) the properties of the $z\sim1.4$ galaxies suggest that the period of rapid chemical evolution  takes place progressively in lower mass systems as the
  universe ages, and thus provides further support for a {\sl downsizing} picture of galaxy
  formation, at least from $z\sim1.4$ to today.
\end{abstract} 
 
\section{Introduction} 
 The starlight that is seen at any particular redshift must be associated with
the production of heavy elements which can then be seen (in stellar
atmospheres or in the gas phase) at all later epochs.  Monitoring how the
chemical content of galaxies changes with cosmic time is thus important to constrain
the history of the global star formation activity in the universe.

Efforts to determine the gas metallicity of star forming galaxies as a
function of cosmic time  have returned oxygen ([O/H]) gas metallicities
for relatively large sample of galaxies up to $z \sim 1$ and a handful
above $z \sim 2$. Most of
these studies have relied on rest-frame optical lines, and in particular on
Pagel's \cite{pagel79} $\rm R_{23}= ({\rm [O\,II]\,}{\lambda\,3727} + {\rm
  [O\,III]\,}{\lambda\lambda\,4959,5007})/ {\rm H}{\beta}$ metallicity
indicator. While relatively easy to measure, $R_{23}$ is
however degenerate with metallicity (as low values of $R_{23}$ may correspond
to very low or high metallicities), and severely affected by dust extinction.
Both problems are solved if the ${\rm H}{\alpha}$ and ${\rm
  [N\,II]\,}{\lambda\,6584}$ line fluxes are also available, since the ${\rm
  H}{\alpha}/{\rm H}{\beta}$ line ratio provides an estimate for the dust
extinction, and the ${\rm [N\,II]\,}{\lambda\,6584}$/${\rm H}{\alpha}$ line
ratio breaks the degeneracy in $\rm R_{23}$.
However, for $z> 0.5$  the ${\rm H}{\alpha}$ and ${\rm
  [N\,II]\,}{\lambda\,6584}$ lines are shifted in the
near-infrared (NIR), and are thus much more challenging to measure than the
lines which appear within the optical window. 
In this  contribution we describe new near-infrared 
spectroscopy enabling us to determine
oxygen abundances for a sample of 30 $0.47<z<0.92$ CFRS and five $z\sim1.4$
galaxies, to investigate how the properties of the star forming gas
evolve in galaxies, from the period of the peak of the star formation
and metal production rates until today.

\section{Near-Infrared Spectroscopy}

30  CFRS galaxies at $0.47<z<0.92$ with absolute B magnitudes
$\rm{M}_{\rm{B,AB}}\la-19.5$  were extracted 
from the 66 objects sample presented in  \cite{lilly03} (see also
\cite{calilly01}) forming an essentially
random sub-sample. They   have
therefore \OIIIa, \Hb, and \OII\, lines measured with the CFHT. 
Near-infrared spectroscopy for  these CFRS galaxies  was
obtained with the ISAAC spectrograph at the VLT and with NIRSPEC at
Keck II in order to measure their \Ha\,  and  \NII\, line fluxes.

 The five  $z\sim 1.4$ galaxies were selected as \OII\,- emitters  at
 $z \sim 1.4$ from three sources:  two galaxies from the CADIS survey \cite{maier03},
 one from the CFRS survey \cite{lilly95II}, and
 two from unpublished analysis of CFHT emission line searches in the CFRS
 22h field \cite{cramlilly99},\cite{tran04}.
For these $z\sim 1.4$
 galaxies we obtained VLT-ISAAC spectra in the J band, to measure the
 \Hb\, and \OIIIa\, fluxes, and in the H-band, to measure \Ha\, and
 upper limits to the [NII] fluxes. 
The data reduction of the NIR spectra is  described in detail in \cite{maier05}.

%##############################################################################
\section{Line ratios}

As shown in \cite{maier05}, we checked that the detected line emission is not
strongly contaminated by the presence of an AGN
using the log(\OIIIa/\Hb) vs.  log(\NII/\Ha) diagnostic diagram. 
In none of the $0.47<z<0.92$ CFRS the line emission is  dominated by an
AGN. Moreover,  as we  show in a
forthcoming paper (\cite{maier05b}), this is also true for the $z\sim1.4$  galaxies.

It should be noted that all five $z\sim1.4$ galaxies have the high \OIIIa/\OII\, and
\OIIIa/\Hb\, line ratios which are typical of the lower luminosity
intermediate-$z$ CADIS 
galaxies (\cite{maier04}) and of the more luminous Lyman Break Galaxies
(LBGs, \cite{pettini01}) at higher $z$. This is  in contrast with the
bright, intermediate-z CFRS galaxies, which show low \OIIIa/\OII\, and
\OIIIa/\Hb\, ratios.

%##############################################################################
\section{The method to determine oxygen abundances}

 Our approach  to determine gas oxygen abundances is based on the models of \cite{kewdop02pap}, who developed a
set of ionisation parameter and oxygen abundance diagnostics based  on the
use of only strong optical emission lines. 
Using the relations given by \cite{kewdop02pap}, we created a model grid of relative
line strengths as a function of three parameters: the extinction
parameter $A_{V}$, the ionisation parameter $q$, and [O/H].

 The method consists in performing a
simultaneous fit to the \OII, \Hb, \OIIIa, \Ha, and \NII\, lines in terms of
extinction parameter $A_{V}$, ionisation parameter $q$, and [O/H].
The measured [OII], H$\beta$, [OIII], H$\alpha$, and [NII] fluxes  are  compared with the theoretical fluxes,
predicted for each of the $\sim 12 \times 10^{6}$ models of our grid (which covers
a large range of $A_{V}$, $q$, and [O/H] values).
For each of the 30 galaxies, the so-derived best fit models provide the oxygen
abundance [O/H], ionisation parameter $q$, and $A_{V}$, and associated uncertainties.
For details of the method see Sect. 3 in \cite{maier05}.

It should be noted that  our philosophy is to treat all galaxies at high and low redshift  in the same way.
Starting with the
observed line fluxes (ratios) and computing oxygen abundances with the
same method for the low and high redshift galaxies      allows us 
to focus on relative effects between the selected
samples in the expectation that these are likely to be much more robust
than atempts to determine very accurate the absolute metallicity. 
Also  the contribution of Sara Ellison in these proceedings
(\cite{ellison}) shows that one must be consistent with the choice of
metallicity calibration, when comparing different samples of galaxies.

%##############################################################################
\section{The metallicity-luminosity relation}

%%%%%%%%%%%%%%%%%%%%%
\subsection{ $0.4<z<1.0$ CFRS galaxies in the  metallicity-luminosity diagram }
%%%%%%%%%%%%%%%%%%%%%

\begin{figure}[t]  
\vspace*{1.25cm}  
\begin{center}
\epsfig{figure=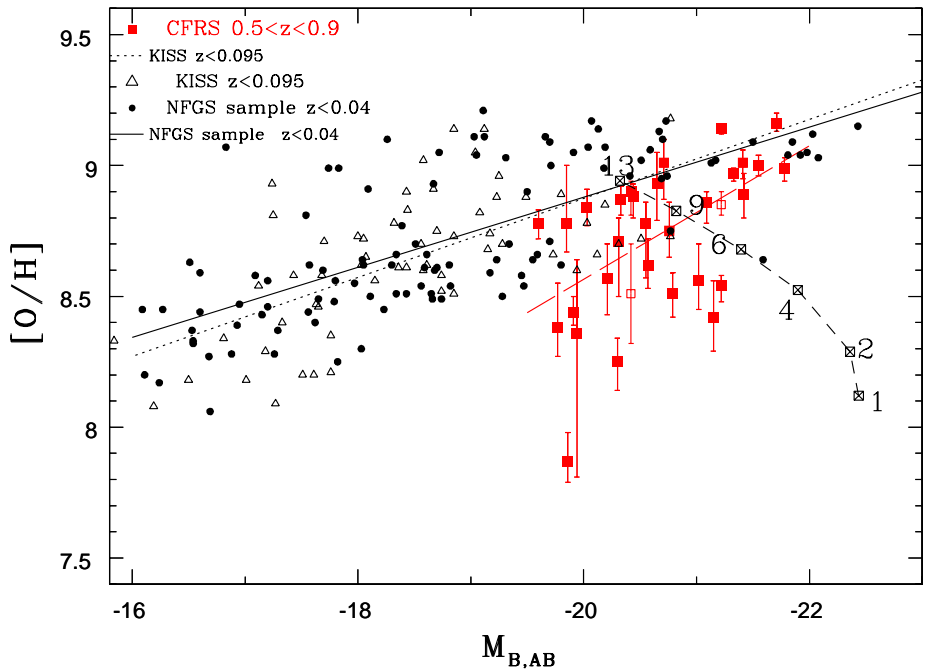,width=12.5cm}  
\end{center}
\vspace*{0.25cm}  
\caption{\label{MB_OH} \footnotesize Oxygen abundance versus $M_{B,AB}$ for
  the 30 CFRS galaxies (filled  squares),  NFGS local galaxies (filled
  circles), and local KISS galaxies (open triangles). The dashed line shows the resulting
  metallicity-luminosity relation  of the 30 CFRS galaxies.
The lower metallicity CFRS
  galaxies are consistent with evolving into NFGS and KISS galaxies
  with slightly (0.5-0.9\,mag) lower luminosities and higher
  metallicities (a factor of  $\sim2$), as shown by an example scenario
 discussed by \cite{maier04}; note that the symbols along the track indicate the
  age of the model galaxy.  
} 
\end{figure}

A metallicity-luminosity relation is observed in the local universe
in the sense that more luminous galaxies
tend to be more metal-rich.  Fig.\,\ref{MB_OH}, shows  $M_{B,AB}$ vs. [O/H]
for the local NFGS (\cite{jansen}) and KISS (\cite{melbsal}) galaxies and the fits to the respective data,
which result in a metallicity-luminosity relation of similar slope and
zero-point for both samples, which we use as local comparison.
The higher metallicity CFRS galaxies overlap with the region of the
diagram occupied by local galaxies of similar luminosity (mass). The lower
metallicity CFRS galaxies ($[O/H]< 8.6$) are more
luminous (massive) than local galaxies with similar [O/H], and more metal-poor than local
galaxies with similar absolute luminosities.

Our new measurements of \Ha\, and \NII\, confirm the low metallicity 
population of galaxies at $M_{B}\sim-21$ found by \cite{lilly03}. It is unlikely
that these  low metallicity galaxies
are the progenitors of today's metal-poor dwarf galaxies  ($M_{B}\sim
-17$), because they would need to fade too much: e.g., in \cite{lilly03} in
Section 4.2  it
was discussed that it is unlikely that CFRS galaxies can fade by more
than 2.5\,mag to the present epoch.  
Moreover, chemical evolution models 
generally produce  ``oblique'' rather than ``horizontal'' tracks:   
the evolution in the metallicity-luminosity
diagram would be from bottom right to upper left, with galaxies evolving from low metallicity and high luminosities
towards higher metallicities and fainter luminosities, as shown by the
model in Fig.\,\ref{MB_OH}.

Comparing the average value of [O/H]
   for the 30 CFRS galaxies to the average value of [O/H] of
  NFGS and KISS local galaxies with similar luminosities in Fig.\,\ref{MB_OH}, we find that  the average change in
  metallicity is  about 0.3\,dex between galaxies at  $z\sim0.7$  and
  $z\sim0$. It is not hard to get a change by a factor of two between
   $z\sim0.7$ and $z\sim0$. For instance,  P\'egase2 models
   \cite{fiocrocca99} discussed by \cite{maier04} can get a  track as
   shown by the dashed curve and symbols in Fig.\,\ref{MB_OH},
   acording to which  the metallicity of the intermediate redshift CFRS galaxies
   may increase by a factor of about 2 by $z\sim0$. 
In this scenario the lower metallicity CFRS
  galaxies may fade by 0.5-0.9\,mag by $z\sim0$, due to decreasing
  levels of star formation, and migrate in the metallicity-luminosity
  diagram  to the region occupied by local galaxies with 
  lower luminosities and higher [O/H].
These are most likely massive spirals. Star formation at $z \sim 1$
likely contributes to building both the disks (\cite{barden05}), and the
bulges (\cite{marci98},\cite{marci99},\cite{marci01}, \cite{marci02}).
%(\cite{marci96}, \cite{marci98}, \cite{marci99}, \cite{marci01}, \cite{marci02}).

%##############################################################################
%  
% Figure 1 
% 
%%%%%%%%%%%%%%%%%%%%%

\subsection{Addition of $z\sim1.4$ and CADIS galaxies to the the metallicity-luminosity diagram}

\begin{figure}[t]  
\vspace*{1.25cm}  
\begin{center}
\epsfig{figure=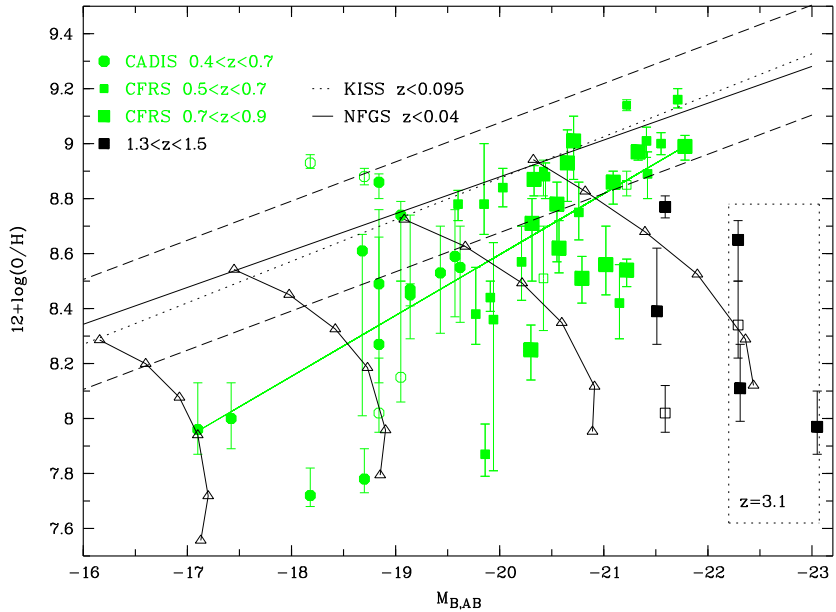,width=12.5cm}  
\end{center}
\vspace*{0.25cm}  
\caption{\label{MB_OHcodeCADIS} \footnotesize Similar to
  Fig.\,\ref{MB_OH},
  but with additional plotted five $z\sim1.4$ galaxies (dark/black filled
  squares) and lower luminosity CADIS
  galaxies (grey/green filled  circles), and LBGs galaxies at $z\sim3.1$. Plotted are also the CFRS
  galaxies, split in two redshift bins: CFRS galaxies at $0.5<z<0.7$
  are plotted as small gray/green filled
  squares, while  CFRS galaxies at $0.7<z<0.9$ are plotted as big
  gray/green filled squares.
The oxygen abundances of the nearby NFGS and KISS galaxies (which are
not plotted explicitly any more) lie between
the dashed lines.
The location of bright galaxies at $z\sim3.1$ is shown as a box
  encompassing the range of  $\rm{M_{B,AB}}$ and [O/H] (without breaking the \R23\, degeneracy)   derived for these
  objects.
Open squares and circles are  alternative (but less probable) oxygen abundance
solutions for some galaxies, as discussed in Section 3.3 and Fig.\,4 of
\cite{maier05}. 
The slope of the metallicity-luminosity
relation at $0.4<z<0.7$  (gray/green solid line),
the lower metallicities  and the ``youthness'' of
the four $z \sim 1.4$ galaxies implied by the
P\'egase2 models   are  a sign of ``downsizing''.  
} 
\end{figure}

To quantify our results for the chemical evolution of
galaxies with cosmic time, we have constructed a grid of chemical evolution
models using P\'egase2, which is described in more
detail in our forthcoming paper \cite{maier05b}.
In order to discuss the evolution of the metal abundance of the star forming gas
in galaxies from $z=0$ to $z\sim1.5$, we compare the P\'egase2 models  with the observed
luminosities and metallicities of the $z\sim1.4$
galaxies in Fig.\ref{MB_OHcodeCADIS}.
 Included in the figure are also the intermediate
redshift CFRS galaxies, and the location of local KISS and NFGS galaxies.

~~~ It is quite noticeable that there is an age-redshift
 relation along a given P\'egase2 track.
For example,   despite the large
scatter, the bright, $M_{B,AB}<-19.5$, $z\sim1.4$ galaxies (dark/black filled squares) appear to be
``younger'' 
 than $0.7<z<0.9$ galaxies (big grey/green filled squares) in the sense that they
 lie towards the beginning of the luminosity-metallicity track.
 The latter  appear in turn
to be on average ``younger'' than most $0.5<z<0.7$ galaxies (small grey/green filled squares)
 which themselves actually overlap
significantly on the diagram with the metallicity-luminosity relation traced
by nearby galaxies.  The tracks of the P\'egase2 models suggest that the
bright star forming $z\sim1.4$ galaxies are likely to evolve into the
population of  less luminous but nonetheless rather massive,
metal-rich galaxies that appear in the $0.5<z<0.9$ galaxy population.
The observation of this age-redshift effect is not surprising, and indeed is reassuring!

%##############################################################################

%##############################################################################

~~~ Adding the CADIS galaxies from \cite{maier04} to the metallicity-luminosity diagram
(filled grey/green circles),  we note  an interesting fact:  the rate
of evolution of a galaxy along its roughly diagonal track appears to depend on
luminosity (mass). 
At the  lookback time of about 6\,Gyrs ($z\sim0.6$)
at which a significant fraction of  more massive $M_{B,AB}\la -20$ galaxies are still on the
zero redshift metallicity-luminosity relation, the lower luminosity (lower
mass) objects at  $M_{B,AB}\ga -19$  sampled by CADIS are already quite
far
from the local metallicity-luminosity.
Moreover, the metallicity-luminosity
relation of the combined CADIS+CFRS $0.4<z<0.7$ sample (solid
line) shows a change in slope compared to the
local metallicity-luminosity relation, in the sense that the offset in metallicity between
$z\sim0$ and $0.4<z<0.7$ is larger at 
lower luminosity than at higher luminosity.
This may be  due to the fact that lower luminosity (mass) galaxies began
their most rapid evolution 
later than high luminosity (mass) galaxies, as
also suggested by the P\'egase2 models and by \cite{kob03}. 
 This is a sign of
downsizing  (see, e.g., \cite{cowie96}, \cite{heavens}, \cite{juneau}, \cite{savaglio}):
 As we look back further in time,
we find signs for departures from the $z=0$ metallicity-luminosity
relation occuring at
progessively higher luminosities (or masses).

%##############################################################################

\section{Conclusions} 
The metallicity of the star forming gas has been measured for 30 CFRS
galaxies with $0.47<z<0.92$ using optical CFHT and near-infrared VLT-ISAAC and
Keck-NIRSPEC spectroscopy.
Using the measurements of five emission lines it was possible to
determine the extinction, oxygen abundances and extinction corrected
star forming rates for these 30 luminous ($M_{B,AB}\la-19.5$) galaxies.

Moreover, the metallicity of the star forming gas has been measured for
five galaxies at $z \sim 1.4$ using new near-infrared VLT-ISAAC spectroscopy
in the H- and J-band,
and already available  measurements of the emission line flux of the
\OII\, line, by which the galaxies were selected.
Using the measurements of \OII, \Hb, \OIIIa, \Ha, and upper limits for
\NII\, it was possible to determine the extinction, oxygen abundances,
and extinction corrected star forming rates for these luminous 
($M_{B,AB}<-21.5$) galaxies.

We  also compared   P\'egase2  chemical evolution
models  with the observed properties of 
galaxies at $0<z<3$.
Our most salient conclusions are:

~~~ 1. The source of gas ionisation in the 30 CFRS galaxies and in the five $z\sim1.4$  galaxies is not
associated with AGN activity, as derived from the log(\OIIIa/\Hb)
versus log(\NII/\Ha) diagnostic diagram. 

~~~ 2. All the $z\sim1.4$ galaxies have the high \OIIIa/\OII\, and
\OIIIa/\Hb\, line ratios which are typical of the lower luminosity
intermediate-$z$ CADIS galaxies and of the more luminous LBGs at higher $z$.
This is in contrast with the
bright, intermediate-$z$ CFRS galaxies, which show low \OIIIa/\OII\, and
\OIIIa/\Hb\, ratios.

~~~ 3. 20 of the 30  CFRS galaxies at  $0.47<z<0.92$ have the higher
metallicities ([O/H]\,$>8.6$) found
locally in galaxies of similar luminosities. However, one third of the
CFRS galaxies have substantially lower metallicities
than local galaxies with similar luminosities and star formation rates.
This is at the upper band of the range  found by \cite{lilly03} for the fraction
of lower metallicities objects, and is due to the fact that we can
account for  the variety of reddening when computing the oxygen abundances. 
We also find that  the average change in
  metallicity is  about 0.3\,dex between the CFRS galaxies  and   local
  galaxies of similar luminosities.

~~~ 4. The evolution of the lower metallicities CFRS galaxies will be
probably  oblique in the metallicity-luminosity diagram: these galaxies
will probably increase their metallicities by about 0.3\,dex and decrease their
luminosities by about 0.5-0.9\,mag, evolving into the region occupied
by today's $z\sim0$ (massive spiral) galaxies. Therefore they are unlikely be the progenitors of
the  metal-poor dwarf  galaxies   seen today.

~~~ 5. There is an age-redshift
 relation along a given P\'egase2 track.
The bright, $M_{B,AB}<-19.5$, $z\sim1.4$ galaxies  appear to be
``younger'' 
 than $0.7<z<0.9$ galaxies, and  the latter  appear in turn
to be on average ``younger'' than most $0.5<z<0.7$ galaxies.
 The tracks of the P\'egase2 models suggest that the
bright star forming $z\sim1.4$ galaxies are likely to evolve into the
population of relatively less luminous but nonetheless rather massive,
metal-rich galaxies that appear in the $0.5<z<0.9$ galaxy population.

~~~ 6. Interestingly,  the rate of evolution of a galaxy along its
track appears to depend on
luminosity (mass).  As we look back  in time,
we find indications for departures from the local metallicity-luminosity relation  occuring at
progessively higher luminosities (or masses). 
The different slope of the metallicity-luminosity relation at
$0.4<z<0.7$ compared to local galaxies, the lower
metallicities and high \OIIIa/\OII\, and
\OIIIa/\Hb\, line ratios of luminous $z\sim1.4$ galaxies 
are a sign of ``downsizing'': the signatures of youth  are seen in more luminous
(and presumably more massive) galaxies at earlier times.

%\acknowledgements{ 
%} 

\vfill 
\end{document}